\documentclass[sigconf]{acmart}
\renewcommand\footnotetextcopyrightpermission[1]{} 
\usepackage{algorithm,algpseudocode}
\usepackage{varwidth}
\usepackage{listings}
\usepackage{xcolor}
\usepackage{graphicx}
\usepackage{makecell}
\usepackage{soul}
\usepackage{enumitem}
\usepackage{titlesec}

\AtBeginDocument{%
  \providecommand\BibTeX{{%
    \normalfont B\kern-0.5em{\scshape i\kern-0.25em b}\kern-0.8em\TeX}}}

\begin{document}

\title{Statically Detecting Buffer Overflow in Cross-language Android Applications Written in Java and C/C++}

\author{Kishanthan~Thangarajah}
\affiliation{%
  \institution{University of Waterloo}
  \city{Waterloo}
  \country{Canada}}
\email{k4thanga@uwaterloo.ca}

\author{Noble~Mathews}
\affiliation{%
  \institution{Indian Institute of Technology Tirupati}
  \city{Tirupati}
  \country{India}}
\email{ch19b023@iittp.ac.in}

\author{Michael Pu}
\affiliation{%
  \institution{University of Waterloo}
  \city{Waterloo}
  \country{Canada}}
\email{michael.pu@uwaterloo.ca}

\author{Meiyappan~Nagappan}
\affiliation{%
  \institution{University of Waterloo}
  \city{Waterloo}
  \country{Canada}}
\email{mei.nagappan@uwaterloo.ca}

\author{Yousra~Aafer}
\affiliation{%
  \institution{University of Waterloo}
  \city{Waterloo}
  \country{Canada}}
\email{yousra.aafer@uwaterloo.ca}

\author{Sridhar~Chimalakonda}
\affiliation{%
  \institution{Indian Institute of Technology Tirupati}
  \city{Tirupati}
  \country{India}}
\email{ch@iittp.ac.in}

\begin{abstract}
  Many applications are being written in more than one language to take advantage of the features that different languages provide such as native code support, improved performance, and language-specific libraries. However, there are few static analysis tools currently available to analyse the source code of such multilingual applications. Existing work on cross-language (Java and C/C++) analysis fails to detect buffer overflow vulnerabilities that are of cross-language nature. In this work, we are addressing how to do cross-language analysis between Java and C/C++. Specifically, we propose an approach to do data flow analysis between Java and C/C++ to detect buffer overflow. We have developed PilaiPidi, a tool that can automatically analyse the data flow in projects written in Java and C/C++. Using our approach, we were able to detect 23 buffer overflow vulnerabilities, which are of cross-language nature, in six different well-known Android applications, and out of these, developers have confirmed 11 vulnerabilities in three applications.
\end{abstract}







\maketitle
\pagestyle{plain}

\section{Introduction}
Applications are being written in more than one language to use functionalities and libraries offered by specific languages. The Android platform allows developers to create applications in more than one language. Android NDK\cite{ratabouil2015android} is a good example where developers can write the applications code in Java and native code using C++ or C. Source code from both languages can be compiled and inter-operate at run-time. In the case of Java and C++, the JNI\cite{gordon1998essential} acts as the bridge for both these languages to inter-operate. 

Developers can reuse already existing modules from those different languages so that the business logic of the implementation can be clearly expressed and the applications can be developed quickly. One of the main use cases that developers make use of in C/C++ is the low-level expressiveness to implement performance-critical components. But multi-language application development has its own challenges. One of the challenges is detecting bugs that are of cross-language nature. This is where a bug is introduced at one layer and its impact materializes in a different layer (e.g., An input to the Java layer causes a segmentation fault in the native layer). 

Buffer overflow, for example, is one of the most common errors that occur in C/C++ where the control on how to access (read/write) a buffer is given to the developers. Without proper checks on how buffers are accessed, the application could crash at run time. In the case of applications written in Java and C/C++, the input data originating from the Java side (source) can end in a buffer access (sink) at the C/C++ side can potentially cause a buffer overflow.

As opposed to single-language-based software systems, cross-language analysis is not an easy task. The major challenge is that we have to normalize the analysis across both languages which may be considerably different and statically cross borders from one language to another. For example, to detect buffer overflow of cross-language origin, we need to precisely track the data flow from Java to C/C++ and integrate the data flow analysis from both languages, as they have different language syntax and semantics. Thus when using multiple languages, there is a requirement to build a single tool that can analyse how the data or control flow from one language to another. 

Existing approaches from Lee et al. \cite{lee2020broadening} and Wei et al. \cite{wei2018jn} on Java and C/C++ cross-language analysis can not detect buffer overflow of cross-language nature due to various reasons (which we explain in detail in Section \ref{lab:motivation}). But in light of source code availability for a program, we propose an orthogonal approach to the work from Wei et al.~\cite{wei2018jn} for detecting buffer overflow that relies on source code static analysis. Our approach is intrinsically more accurate, as it leverages richer and more precise information compared to binary code. We use an abstract representation of both language pairs using a common XML-based AST representation and then perform a forward program slicing to precisely track the data flow from Java to C/C++ that ends in a potential buffer overflow.

\textbf{The main contributions of this work are as follows.} \hfil
\begin{enumerate}
\item We present an approach to do cross-language static data flow analysis. The data flow analysis solution is based on an XML-based AST representation of source code and forward program slicing.
\item We have developed PilaiPidi, a cross-language static data flow analyser. To the best of our knowledge, PilaiPidi is the first solution that addresses the problem of cross-language analysis to detect buffer overflow.
\item We have evaluated PilaiPidi on real Android applications and we were also able to detect and confirm 11 new buffer overflow vulnerabilities from three different Android applications using our approach.
\end{enumerate}

\begin{figure}
\begin{lstlisting}
01. // YuvOperator.java source file
02. public class YuvOperator {    
03.     native static void jniRotate(ByteBuf handler);
04.
05.     void rotate(byte[] yuv, int width) {
06.         handler = jniStoreYuvData(yuv, width);
07.         jniRotate(handler);
08.     }
09. }
\end{lstlisting}
\caption[fig5]{Java native class}
\label{fig:exmaple_java_native_class}
\end{figure}
\raggedbottom

\section{Background and Threat Model}
\textbf{Cross Language Invocation}: Applications using multiple programming languages are implemented with the support of foreign function interfaces (FFI) offered by those languages. Different languages have different ways to provide external function invocation support in their language syntax and semantics. JNI is the FFI offered by Java to invoke C/C++ functionalities within Java. Some languages do support the invocation from the target language to the source language making the invocation happen in both directions. 

Java, for example, uses \textit{native} keyword to mark functions that are implemented at the native level and invoke them using the JNI functionality. Java methods can also be invoked directly from C/C++ language using the JNI library functions available at the native level.  Foreign function registration (binding) is supported either statically or dynamically in Java. Static binding is a default approach where native functions with JNI are registered during compile time with the specific function notation. With dynamic binding, the user has to make sure to load the required JNI library during the linking time and use the \textit{onLoad} functionality provided by the JNI run time.

\textbf{Buffer overflow}: A buffer is a continuous section of allocated memory of a defined type such as integers, characters, etc. A buffer overflow  occurs when a program in execution tries to insert more data than what the buffer can hold. Writing and reading outside of the buffer bound can cause issues like corrupting the data, crashing the program in execution, or even opening the door for the execution of malicious code.

Buffer-bound checks are not carried out by most of the memory manipulation functionalities offered by C/C++. It is up to the programmer to check for the buffer bounds and write code accordingly. But most of the time, the code written by the programmers assumes wrong size values or wrong size calculations when checking for buffer-bound access. Both programmers' incorrect assumptions and access to direct memory manipulation exposed by the functions are considered the root cause of buffer overflow.

\textbf{Threat Model}: In Android applications, the native functions  written in C/C++ will be packaged as library files which will be loaded at run time for execution. For buffer overflow occurring on the native side, the input can originate from the Java side. A malicious user who discovers such data flow paths can send arbitrary input data to the application, which overflows the allocated memory size for the buffer on the native side. By exploiting this input, the attackers use buffer overflow to corrupt the execution stack of an application. For this work, we are focusing on such data flow paths originating in the Java side, where the user input is captured externally and ends in the native C/C++ side which leads to a buffer overflow. We then exploit such paths by sending arbitrary data as input values to the Java methods and observe whether such input values cause a buffer overflow error on the native side.

\begin{figure}
\begin{lstlisting}
01. // JniYuvOperator.cpp source file
02. JNIEXPORT
03. void JNICALL Java_YuvOperator_jniRotate
04.     (JNIEnv *env, jobject obj,  jobject handle) {
05.     JniYuvOperator *yuvOperator = 
06.         env->GetDirectBufferAddress(handle); 
07.     char *yuv = yuvOperator->_yuvData;
08.     int width = yuvOperator->_width;
09.     std::vector<unsigned char> yuvCopy(yuv); 
10.     int n = 0;
11.     for (int i = 0; i < width; i++) {
12.        yuv[n++] = yuvCopy[width * i];
13.     } 
14. }
\end{lstlisting}
\caption[fig5]{Native C++ source for the class in Figure \ref{fig:exmaple_java_native_class}}
\label{fig:exmaple_cpp_native_header}
\end{figure}
\raggedbottom

\begin{figure*}
\includegraphics[width=\textwidth]{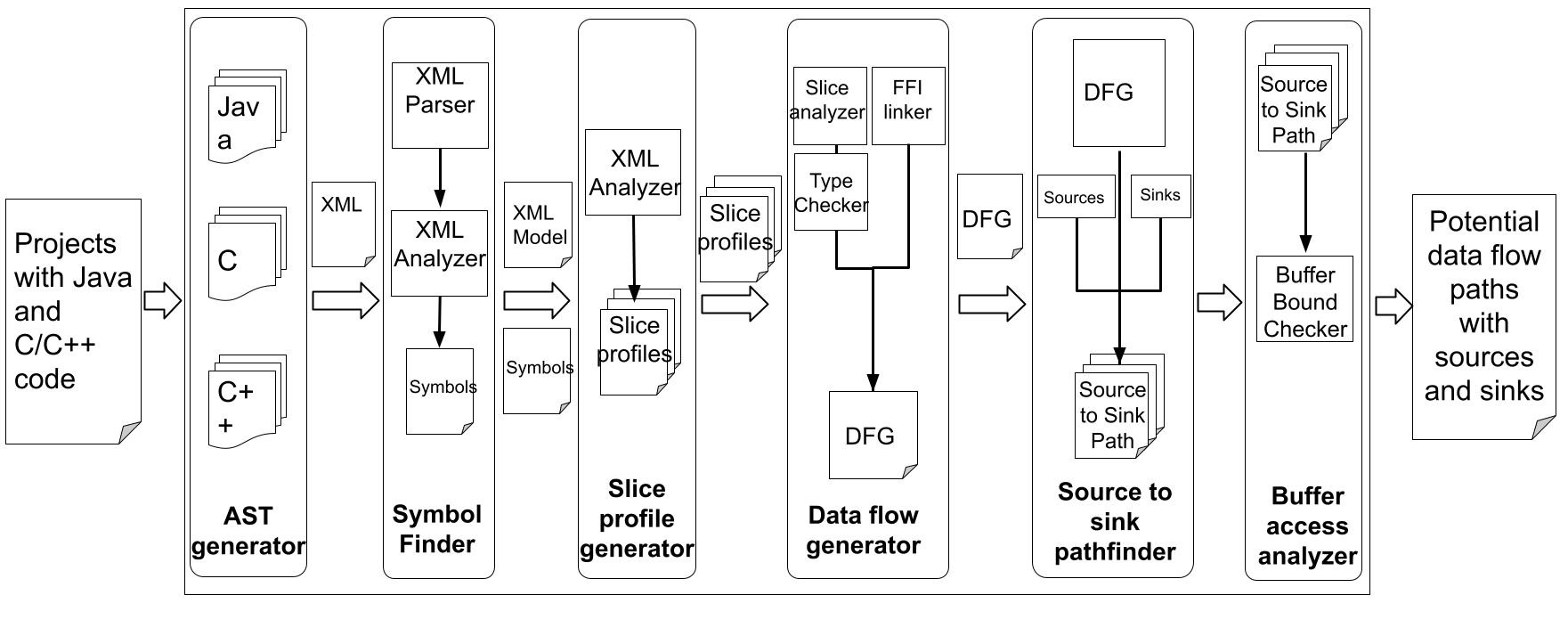}
\caption[fig6]{Overview of PilaiPidi}
\label{fig:overview_of_pilaipidi}
\end{figure*}
\raggedbottom
 
\section{Motivation} \label{lab:motivation}
Let us look at an example Java-to-C/C++ cross-language inter-operation from Figure \ref{fig:exmaple_java_native_class} and Figure \ref{fig:exmaple_cpp_native_header}. The example in Figure \ref{fig:exmaple_java_native_class} and \ref{fig:exmaple_cpp_native_header} is an oversimplified version of a real buffer overflow which we have identified in an open-source Android application \footnote{We have reported the issue to the developers. The flow has been acknowledged and fixed recently.} \cite{pilaipidifixes}.  
 
The flow starts from line 05 in Figure \ref{fig:exmaple_java_native_class} when the method \textit{rotate} is invoked which in turn calls the JNI method \textit{jniRoate} at line 07. This method takes the reference pointing to the \textit{handler} instance as the argument and the invocation continues into the native C++ side at line 04 in Figure \ref{fig:exmaple_cpp_native_header}. The buffer \textit{yuv}, which is passed from the Java side is acquired at line 07 and the variable \textit{width} is accessed at line 08 which. The buffer \textit{yuv} is  accessed at line 12 within a loop and the loop condition uses \textit{width} at line 11 for the termination. 

This example shows that a buffer \textit{yuv} which was passed from the Java side is accessed on the C++ side, and the buffer is accessed using index variable \textit{width}. In this example, a potential buffer overflow read can occur at line 12, on the right-hand side of the statement, when the \textit{yuvCopy} buffer is accessed since the buffer size is not explicitly checked before the access. 

Existing approaches \cite{lee2020broadening}, \cite{wei2018jn} on cross-language analysis for Java and C/C++ proceed by leveraging tools tailored towards each individual language and combining their results. 

The work from Lee et al. \cite{lee2020broadening}, uses semantic summary-based analysis and looks at the issues that can arise due to interoperability, such as having wrong foreign function calls or mishandling of Java exceptions. This work cannot identify cross-language buffer overflow vulnerabilities, as in the example above, since it is not specifically concerned with data flows that causes buffer overflow across the two languages.

The work from Wei et al. \cite{wei2018jn} tackles data flow tracking via a heap summary-based bottom-up approach across Java and C/C++ on binary code. However, we note that the approach is intrinsically limited and cannot detect buffer overflows: (1) it cannot detect buffer-relevant sinks such as reads or writes and (2) cannot identify code properties characterizing buffer overflows (e.g., does resolve the buffer sizes and detect the cases where a Java-level source buffer overflows a native-level destination buffer.

In this work, we propose an orthogonal approach for detecting cross-language errors, tailored toward buffer overflow. Our approach works at the source code level -- thus, it is orthogonal in nature to existing binary-level solutions. We unify the detection process of buffer overflow across the Java and native layers with our proposed abstract representation of the two different languages. 

Specifically, we use an XML-based abstract representation of the source code for both Java and native layers and then perform forward program slicing to build the data flow graph for the whole application. With the analysis we perform with program slices, we gain access to context-sensitive and precise information such as buffer access locations and how to accurately model safeguards around buffer access which we use to filter false positive sink locations. The abstract XML representation unifies the buffer detection process across the two layers. In the next section, we will present the details of our approach. 

\section{Cross-language Buffer Overflow Detection with PilaiPidi} \label{lab:design_and_developement}

We first convert both Java and C/C++ to a common format, which preserves all the original information from the source file. Specifically, in our design, we choose an Abstract Syntax Tree (AST) representation of the source code. The AST representation is captured in an XML format. 

We then do data flow analysis by using forward source slicing. For this, we built a program slice generator that works on top of XML-based AST output and created slice profiles. We then combine both XML and slices from both languages to analyse and build data flow paths and then check for potential sources and sinks. This also involves finding the cross-language link between two different languages. Finally, we check for potential buffer access issues from the identified data flow paths. The overview of the PilaiPidi architecture is given in Figure \ref{fig:overview_of_pilaipidi}. It has six major phases namely, AST representation generator, symbol finder, slice profile generator, data flow generator, source-to-sink pathfinder, and buffer access analyser. 

\subsection{AST representation generator}
In this phase, we want to extract the source code from different languages to a common AST representation. 
We used srcML \cite{collard2013srcml}, which is an open-source tool, to convert source code from both Java and C/C++ to an AST representation that is saved in an XML format. The AST representation provides a platform to extract, analyse and convert source code. 
As it preserves all the source information, including comments and position, it gives us enough flexibility to analyse, both syntactically and semantically, the source without losing any information. 

This phase generates the same XML format for all the languages at the AST representation level. Some of the common constructs are class, function, function\_parameters, function\_block, statements, loops, conditions, etc. These constructs are commonly represented as XML tags, attributes, and elements for all the supported languages. With this advantage, we were able to reuse the same data flow analysis solution for both Java and C/C++ in our design. To get an understanding of how both Java and C/C++ sources are represented in XML format, we created an online appendix with some examples.\footnotemark[1]{}
\footnotetext[1]{https://figshare.com/s/8b8f3f1b3b98191e82de}


\subsection{Symbol finder}
To do proper type checking, we have to have type information of all the structures, classes, variables, and methods (i.e type information of the method definition). Symbol finder is a separate phase that goes through all the source files in the whole project and finds type information. The reason to separate out the symbol-finding phase is to avoid and handle scenarios on unresolved types at the time of the slice-building phase. Also, this makes the architecture clean with separate phases with relevant components so that any changes to one phase do not affect the other phase(s). 

The Symbol finder first parses the XML to generate a Java model and then analyses the top-level constructs. The possible top-level constructs that the symbol finder looks for are Structures, Classes, and Functions. For each of these constructs, the symbol finder starts by capturing the name and its type and then analysing the content body of them. The output XML represents source code using a collection of XML elements. For example, the C++ language constructs such as structures, classes, and functions are represented as top-level XML elements, and their content is represented as sub-elements \cite{srcmldoc}. 

A class in C++ can have public, private, or protected block contents and each of these blocks can have fields such as variable declarations and function declarations. Similarly, a function declaration can have one or more statements such as variable declarations, conditional statements, looping statements, etc. These constructs are captured as top-level constructs and their body in a hierarchical XML representation. The XML parser converts the XML format to a Java model, which we call the XML model from now onwards. 

At the end of this phase, we will have all the symbol information captured. A symbol has three instance variables: name, type, and parent. The name and type are the name and type of the symbol respectively, while the parent is the parent symbol of the current symbol, which is captured when the symbol finder goes through fields and functions of constructs like structs and classes. 

The collected type symbol data is passed into the next phase, which is the slice generator, and used when resolving types for unresolved types while building slice profiles. This makes sure that all the slice profiles have their type name captured which will be used to validate during the data flow generation phase.

\subsection{Slice profile generator}
We rebuilt a forward-slicing tool on top of the XML model. Our solution is inspired by the srcSlice tool \cite{newman2016srcslice}. We initially tried to use the srcSlice tool, but due to some technical limitations \footnotemark[1]{}, and the need for extending to support our requirements such as capturing and propagating type symbol information, we implemented our own slice generator on top of XML output. The slice generator analyses the XML model and produces a dictionary of slice profiles for all the identifiers, i.e the variables, found in all the source files. For each identifier in the source, a slice profile is produced and a slice profile consists of a list of elements as below.
\footnotetext[1]{https://figshare.com/s/dafa4e162796f2e50dc4}
\begin{itemize}
\item fileName - source file name of the slice variable
\item functionName - function name of the slice variable
\item varName - identifier name of the slice variable
\item typeName - type name of the slice variable
\item definedPosition - line number information where the identifier is defined
\item usedPositions - list of line number information where the identifier is used
\item dependentVars - list of data-dependent variables of this identifier
\item cFunctions - list of functions invoked using this identifier
\end{itemize}

As an example, if we consider the source code given in Figure \ref{fig:exmaple_java_native_class}, there are two functions, and the first function, \textbf{\textit{jniRotate}}, has one variable which is the function argument and this function does not have a body. The second function, \textbf{\textit{rotate}}, has three variables, where two are function arguments and the third one is a function local variable \textit{handler}. There will be four slice profiles for the four variables found from the two functions on the Java side. Similarly, for the example in Figure \ref{fig:exmaple_cpp_native_header}, for each of the variables in this function, we will have a separate slice profile generated.

During the slice profile generation, we also initialize the value for each profile. The size (in the case of a buffer) or the value (in the case of a variable), will be captured when each slice profile is created and the value will be included with their respective slice profiles. The buffer size can be captured when there is an explicit buffer size given when the buffer is defined. For variables/identifiers, the size can be identified when the variable is first defined. As we are targeting only the buffer access-related error, we limit the type of the values to integers only. For all other variables/identifiers, the value will be initialized to null as we will not consider them when checking buffer-bound conditions. The value captured for each profile is not the final value as they can be later updated using an assignment statement. Updating the values is explained in the data flow generation phase.

We then use the slice profiles generated by the generator and then collectively analyse all the profiles to build a data flow graph in the next phase. Our idea is that, when we successfully analyse all the slice profiles and build a graph using the information, we should be able to have one or more source nodes and one or more sink nodes, where sources will be in one language and the sinks will possibly be in another language. But the challenge would be to correctly link two languages at their foreign function interface (FFI) layer.

\subsection{Data flow generator}
We now combine both the XML model and slice generator outputs and build a solution that can generate data flow. The XML model for each source file and slice profiles generated from the previous phases are used as the input for this phase and both of them are fed into our Algorithm \ref{fig:data_flow_analyse_alogorithm} to analyze. 


Let us look at Algorithm \ref{fig:data_flow_analyse_alogorithm} from line 4, where we initially start to analyse each source file and its slice profiles. We start randomly as we are trying to build a complete data flow graph with variables and function names. We try to construct each node in the graph using four parameters. The name of the variable, enclosing function name, enclosing source file name, and defined position of the variable are the four parameters that we use to uniquely identify each node in the graph. 

The next step is to start analysing the slice profiles generated for a source code file at line 06. As explained previously, each slice profile consists of multiple elements and we are currently interested in two such elements. First is the cfunctions list, which describes the functions that are invoked using the data from the sliced variable, and second is the dependent variable list. We start with this cfunction list and for each of the cfunction along with the argument position, we then try to find the slice profile from the global list \textit{L}. 

When trying to find the correct slice profile, we further do validations to avoid possible false positive results such as matching the number of arguments and their type. This validation logic can further be improved with more robust rules such as cross-verifying the source file name in the imports list, validating the function type signature as a whole, etc. When we find the possible source slice, we then repeat the same steps from line 8. When there are no cfunctions in the profile, we evaluate the other condition in line 16. 

If the enclosing function of the current slice profile is native, then we have reached an FFI on the Java side. Now we try to map the enclosing function and its signature to a possible JNI function in C++ source code. This is explained in \ref{crosslanglink} section. By linking and then finding the correct slice profile, we then resume analysing the profile from line 8. We then continue and check for any dependent variables~(dvars) from the current slice profile. 

If there are dvars found, then we try to get the slice profile of that dvar and then continue from line 8. This procedure is repeated for all the files and their slice profiles.

\begin{algorithm}[t]
\caption[fig6]{Algorithm to analyse slice profiles of a given project}
\label{fig:data_flow_analyse_alogorithm}
    \begin{varwidth}[t]{\textwidth}        
       \begin{algorithmic}[1]
  \State $L$: a list of slices for all the source files
  \State $G$: a data flow graph to capture nodes and edges
    \Procedure{Analyse Slices}{$L$, $G$}
        \For{each source file $i$ in $L_i$}
            \State $P_{l_i} \gets$ list of slice profiles in $L_i$
            \For{each slice profile $j$ in $P_{l_i}$}
                \State $encl\_func \gets$ get enclosing function 
                \For{each cfunc $k$ in $P_{l_i}$ } \label{spanalyse}
                    \State $arg\_pos_{k} \gets$ argument positional index
                    \State $arg\_type_{k} \gets$ argument type
                    \State $cfunc\_name_{k} \gets$ cfunction name
                    \State $SP_{p_k} \gets$ find in $L$ using $arg\_type$ \& $arg\_pos$
                    \If{$SP_{p_k}$ found}
                        \State $G \gets$ add edge from $encl\_func$ to cfunc
                        \State \textbf{go to} \ref{spanalyse}
                    \ElsIf{$encl\_func$ is a FFI}
                        \State $SP_{p_k} \gets$ find using $arg\_pos$ \& $cfunc\_name$
                        \State $G \gets$ add edge from $encl\_func$ to FFI func
                        \State \textbf{go to} \ref{spanalyse}
                    \EndIf
                \EndFor
                \For{each dvars $m$ in $P_{l_i}$ }
                    \State $SP_{p_m} \gets$ find slice profile in $L$ using dvar 
                    \If{$SP_{p_m}$ found}
                        \State \textbf{go to} \ref{spanalyse}
                    \EndIf
                \EndFor
            \EndFor
        \EndFor
    \EndProcedure
  \end{algorithmic}
    \end{varwidth}\quad\quad\quad              
\end{algorithm}
\raggedbottom

\subsubsection{\textbf{FFI linking}} \label{crosslanglink}
Mapping from Java JNI native methods to C/C++ JNI functions was carried out using the conventions followed by JNI headers. We use \textit{Java\_packageName\_className\_methodName} pattern to find the C++  file that contains the JNI function for a Java native method. From the identified C++ source files and their slice profiles, we then try to get the correct slice profile of interest by mapping the argument name and argument positional index. 

When a possible slice profile is found, we resume processing from line 8 in Algorithm \ref{fig:data_flow_analyse_alogorithm}. From the motivating example in Figure \ref{fig:exmaple_cpp_native_header}, the name \textit{Java\_YuvOperator\_jniRotate} will be identified using the class name \textit{YuvOperator} and method name \textit{jniRotate}.

\subsubsection{\textbf{Updating the potential value of each profile}}
This step in this phase is required for the buffer access analysis phase. When generating the data flow paths by analysing slice profiles, we also update the value for each slice profile. As mentioned during the slice profile generation phase, the size (in the case of a buffer) or the value (in the case of a variable), will be captured when each slice profile is created and the value will be included with their respective slice profiles. 

Though a variable can be defined in one location, it can be updated or assigned another value later in the program execution flow. For such scenarios, we update the value to a new value, which will either refer to a new value or to another variable's value as a reference. The data flow generator identifies statements that are buffer read or write and updates the value of the corresponding slice profile. This is carried out when analysing the data-dependent variables of the slice variable.

With the approach we followed above, we have built a cross-language analysis solution with the complete data flow graph generated. The generated data flow graph nodes from the motivating example are given in the online appendix.\footnotemark[1]{}
\footnotetext[1]{https://figshare.com/s/8b8f3f1b3b98191e82de}. At the end of this phase, we will have a data flow graph that has captured all the nodes and edges.

\subsection{Source to sink pathfinder} \label{subsec_sourcesinks}
This phase takes the data flow graph as the input, identifies the potential data flow paths, and outputs them as a list of source-to-sink paths. We now explain how we identify potential sources and sinks in this phase.

\begin{table}
\begin{tabular}{|l|l|l|l|l|}
\hline
\textbf{Algorithm}  & \textbf{Accuracy} & \textbf{Precision} & \textbf{Recall} & \textbf{F1} \\ \hline
Naive Bayes         & 68                        & 71                         & 68                      & 67                        \\ \hline
Random Forest       & 93                        & 93                         & 93                      & 93                        \\ \hline
SVM                 & 93                        & 93                         & 93                      & 93                        \\ \hline
Logistic Regression & 95                        & 96                         & 95                      & 95                        \\ \hline
\textbf{XGBoost}    & \textbf{95}               & \textbf{96}                & \textbf{95}             & \textbf{95}               \\ \hline
\end{tabular}
\caption{Performance of the models on the test dataset}
\label{tab:susi_performace}
\end{table}
\raggedbottom

From the analysis perspective of PilaiPidi, a source is an entry point to the application which triggers an invocation. Typically an entry point in an Android application is external facing which takes user input. We describe below how we identify sources with our approach. We first describe the classifier we used to generate the list of source functions in the Android Framework.

\textbf{Classification of Android SDK APIs}: Susi \cite{arzt2013susi} is a machine-learning-based technique for locating sources and sinks in the source code of any Android SDK. Susi finds sources and sinks in the SDK based on a training set of hand-annotated sources and sinks. Susi further categorizes sources as unique identifiers, location information, etc., and sinks as network, file, etc., to give more fine-grained information. 

Susi extracts several semantic and syntactic characteristics from the Android API methods. The collected features are then used to train a machine-learning algorithm on a small subset of manually categorized Android API methods. Susi then uses this model to categorize a number of previously undiscovered Android API methods. Susi's categorization process consists of two parts. Susi determines if an Android method is a source, sink, or neither-nor in the first stage. The second classification model performs semantic category classification on techniques that the first classifies as either a source or a sink method. Methods labeled as neither-nor by the first classifier are ignored. Susi determines the most likely semantic category for an unclassified source or sink. Each method is assigned exactly a single semantic category.

\textbf{Classification approach}: For our study, we utilize the features that are generated by Susi to build a source/sink classifier. Susi was used to building a training dataset for our classifier. The methods from Android 17 SDK are annotated either as source/sink/neither-nor using the Susi classifier. A total of 8272 methods from Android 17 SDK are annotated using Susi, out of which 6621 methods are used to train the classifier, and 1651 methods are used as the validation set. 155 methods that are hand-annotated are used as the testing dataset. 

The machine learning algorithms used to develop the classification model are SVM, Random Forest, Naive Bayes, Logistic Regression, and XGBBoost. Table \ref{tab:susi_performace} shows the classification results for the test dataset. From this table, it can be seen that Logistic Regression and XGBoost perform better than the other algorithms. We then used XGBoost to generate the classification labels for the methods present in the Android 30 SDK. From the classification generated, we used the list of functions that are classified as sources from the Android 30 SDK for the next step.

\textbf{Source functions finding approach}: We take the classified source functions in the Android 30 SDK from the previous step to identify sources in the source code. This first involved identifying data paths where data is ingested on the Java side of the program and then passed to the C++ side. Then, we checked if there were any function calls along this path that matched the name of a function on the list of source functions we identified with the classifier. However, with this approach where we simply match function names, we risk counting false positives, where there may be a function that coincidentally matches the name of a function on our source list, but it is not actually a source function in the Android SDK.

We try to reduce the false positives by adding two additional filters. First, we check if the function call is a local call since calls to Android SDK functions are not local calls as they have to call the method of an external class.

Secondly, we check if the number of parameters that the function is called within the data flow path matches the number of parameters of the Android 30 SDK function in our source list. Our tool was then used to analyze open-source projects that were using JNI to collect data on the number of potentially vulnerable sources in the data flow paths of these projects.

\subsubsection{\textbf{Sinks}} \label{soruceandsinks}
To identify the sinks portion of the source-to-sink pathfinder, we capture sinks related to five categories. The source-to-sink pathfinder is provided with a list of functions that correspond to potential sinks in C/C++. A sink function, in this context, is a function that takes ownership of an input pointer and is responsible for the allocated memory. To improve on our existing list of sink functions, we curated an updated list of potential C/C++ sink functions by reviewing existing literature. 

We describe different types of categories related to curating functions from Sun et al. \cite{Ding_2014} in Table \ref{tab:sources_categories_summary}. The work from Sun et al. \cite{Ding_2014} identifies various patterns for common categories of buffer overflow-sensitive sinks. These common categories are the first four categories given in Table \ref{tab:sources_categories_summary}. Apart from these four categories, we also capture locations where buffer read and write occurs as potential sinks as the fifth category. This is a critical part of our approach which differentiate our work from existing related work \cite{lee2020broadening} \cite{wei2018jn}. We look for places where buffer read or write happens and check whether a such location is found in data flow paths constructed from the data flow graph. When such buffer access locations are found on the data flow path, we mark those as potential sinks. 

Below we explain in detail the identified five categories of buffer overflow-sensitive sinks.

\begin{table}[t]
    \begin{center}
        \setlength{\tabcolsep}{10pt} 
        \renewcommand{\arraystretch}{1.5} 
        \begin{tabular}{| m{8em}| m{14em}|}
            \hline
            \textbf{Category} & \textbf{Example Functions} \\ [0.5ex] 
            \hline
            \#1 - Input & \verb_scanf_ and its family of functions, \verb_fread_, \verb_getc_, \verb_gets_\\ 
            \hline
            \#2 - Memory & \verb_memcpy_, \verb_memmove_, \verb_strcat_ and its family of functions \\
            \hline
            \#3 - Output & \verb_printf_ family of functions, \verb_putc_ and related functions, other output functions defined in \verb_stdio.h_ \\
            \hline
            \#4 - Utility & \verb_realpath_, \verb_getwd_, \verb_getopt_, and \verb_getpass_ \\
            \hline
            \#5 - Buffer Access & Any type of buffer (array) read or write \\
            \hline
        \end{tabular}
    \end{center}
    \caption{Sinks categories with examples}
    \label{tab:sources_categories_summary}
\end{table}
\raggedbottom

\textbf{Category \#1: Input Functions}: The first pattern involves input functions that define or update a destination buffer. These include C stream input functions from \verb|<stdio.h>| and C++ input functions inherited from \verb|istream|. The main concern with these input functions is the potential for them to write past the end of their destination buffer if the length of the buffer is shorter than the input data. Additionally, if the input does not match the type of the input format string given, then the input data could be incorrectly parsed into the destination buffer. For example, reading an integer into a string buffer could result in a buffer overflow since the string would not be properly null-terminated.

\textbf{Category \#2: Memory Functions}: The second pattern matches functions that copy or move the contents of a block of memory to another block. If the size of the destination memory block is smaller than the size of the source block, then data past the end of the destination block may be overwritten. Additionally, if parameters to the moving/copying function contain an off-by-one error, it could corrupt the data, such as in the case where the null terminator of a string is not accounted for in its length and it is not copied over to the destination block. 

A concrete example of this is if the \verb|strncpy| function is used to copy the string \verb|"abcde"| into a char array of size 3 with a maximum character limit set to \verb|3|. At first glance, there does not seem to be a problem since the destination array has a size of 3 and we are limiting the number of characters being copied over to 3. However, the destination array would be filled with the characters \verb_[`a', `b', `c']_, which is not null-terminated. This could cause problems if this array is used as a string in the future and assumed to be properly null-terminated.

\textbf{Category \#3: Output Functions}: The third pattern includes C stream output functions that are declared in \verb|<stdio.h>| which take in a format string whose output is dependent on its parameters. If the format string does not match the data types of the parameters passed in, it could result in a buffer overflow. An example of this could be using the format string \verb|"%s"|, but passing in an integer as the parameter. Since the integer likely does not contain a byte that is equivalent to a null terminator, the program will continue reading past the end of the variable.

\textbf{Category \#4: Utility Functions}: This category includes functions that are considered utility functions in the C/C++ library that could potentially result in a buffer overflow. Generally, they write to some destination buffer and the end of which can easily be overwritten if the buffer is not large enough. Examples of these include the \verb|realpath()| and \verb|getwd()| functions, which write a path to a buffer that is passed into the function as a pointer. Since it is not known at compile time exactly how large the path that is returned will be, if the buffer allocated is too small (less than \verb|PATH_MAX|), it may be overwritten~\cite{Lhee_2003}.

\textbf{Category \#5: Buffer Access}: This is a special category where, along the data flow path, we capture all the nodes that contain a buffer read or write operations. If we consider the same example from Figure \ref{fig:exmaple_cpp_native_header}, we can see indexed-based buffer access (read) at line 12 on the right-hand side. We get this information from the XML model of the source code and our tool identifies such nodes in the data flow path by checking for index-based buffer access expressions in a given statement. Table \ref{tab:sources_categories_summary} describes the five categories with examples for each of them.

\textbf{Source-to-sink path finding approach}:
The source-to-sink path-finding component makes use of the data flow graph generated by the data flow generator and converts the problem into a graph analysis problem. The data flow graph output from the previous step is a complete graph containing edges and nodes. Hence finding paths with sources and sinks is modeled as finding the shortest path from the source node, which is on the Java side, and the sink node, which is on the C/C++ side. For each source node, we try to find all the paths to the target sink node. 

When we construct the data flow for the  example in Figure \ref{fig:exmaple_java_native_class} and Figure \ref{fig:exmaple_cpp_native_header}, we find that the buffer \textit{yuv} is the start of the data flow path at line 5 in Figure \ref{fig:exmaple_java_native_class}. The data flow path will correctly capture that the source node in the path will be the \textit{yuv} from line 5 in Figure \ref{fig:exmaple_java_native_class} and sink will be the \textit{yuvCopy} buffer read operation at line 12 in Figure \ref{fig:exmaple_cpp_native_header} and it will be marked as a potential venerable path. The identified potential data flow path with buffer read is given in the online appendix examples.\footnotemark[1]{}
\footnotetext[1]{https://figshare.com/s/8b8f3f1b3b98191e82de}

\subsection{Buffer access analyzer}
This is the final phase which checks whether there exists a bound check along the data flow path. The ability to do such checks differentiates our work from the other related work \cite{lee2020broadening}, \cite{wei2018jn}, where they cannot do such checks. This is explained in Section \ref{lab:motivation}. 

The input to this phase is the list of source-to-sink paths (data flow paths) from the previous phase. From the sink node of each path, we try to see whether the buffer size that is being used with the buffer access has already been checked with a bound size. If we find a buffer size that has been checked to be within  bounds, we filter that path from the current list and continue to the next path. 

If we consider our motivating example, at line 12 in Figure \ref{fig:exmaple_cpp_native_header}, we can see that there is a buffer read on the right-hand side of the assignment statement. When we reach this node in this phase, we check whether the size of buffer \textit{yuvCopy} is less than the access location. Since the buffer accessed index is related to variable \textit{i}, we compare the value of \textit{i} which is checked against the value of \textit{width}, and the buffer size will be checked against the value of \textit{width} anywhere along the data flow path from the source. 

Currently, in our approach, we have limited these bounds check to just the C/C++ side. In theory, we should be able to expand these bounds check to the Java side too since the data flow graph is in one format.

In our example, there are no bounds checks for \textit{width} and \textit{i} against the size of the buffer as can be seen in Figure 2. Hence, we flag this path as a potentially vulnerable path. If there was a bound check, then to the extent that we can statically check, we see if the access variable is checked against the size of the buffer \textit{yuvCopy}, and we flag this path as a potentially vulnerable path. If there was a bound check, then to the extent that we can statically check, we see if the access variable is checked against the size of the buffer.

The problem of checking the value of buffer bounds can actually be thought of as a value analysis problem, where we check for possible values of a variable that is used as the index in reading from or writing to a buffer. As explained previously, for each slice profile, the slice generator phase embeds the value of each variable, and the data flow generator updates it if it is used or assigned to another variable. By doing reverse tracking of values on variable slice profiles from a specific point, such as buffer read and write, we were able to find buffer size value and buffer index access value.

\begin{table*}[h]
  \centering
  \begin{tabular}{|c|c|c|c|c|c|c|}
    \hline
    & \textbf{Sipdroid} &	\textbf{camerakit} &	\textbf{mupen64plus-ae} &	\textbf{AsciiCam} & \textbf{wonderdroid-x} & \textbf{frozenbubble} \\
    \hline
    grep sinks & 104 & 23 & 3 & 0 & 211 & 725 \\
    \hline
    Flawfinder & 32 & 10 & 3 & 0 & 209 & 432 \\
    \hline
    Cppcheck & 448 & 61 & 18 & 9 & 90 & 740 \\
    \hline
    JN-SAF & Timed-out & Timed-out & Timed-out & Timed-out & Timed-out & Timed-out \\
    \hline
    C-Summary & N/D & N/D & N/D & N/D & N/D & N/D \\
    \hline
    \begin{tabular}{@{}c@{}}PilaiPidi\\ warnings\end{tabular} & 19 & 6 & 4 & 5 & 36 & 22\\
    \hline
     \begin{tabular}{@{}c@{}c@{}c@{}}PilaiPidi\\ warnings exploited\\ by us with citation\\ of the pull request\end{tabular} & 4 \cite{sibdroidfixes} & 6 \cite{camarakitfixes} & 1 \cite{mupen64plusfixes} & 5 \cite{asciicamfixes} & 2 \cite{wonderdroidfixes} & 5 \cite{frozenbubblefixes} \\
    \hline
    Precision \% & 21.1\% & 100\% & 25\% & 100\% & 5.6\% & 22.7\% \\
    \hline
    \begin{tabular}{@{}c@{}c@{}}PilaiPidi\\ warnings confirmed\\ by developers\end{tabular} & 4 & 6 & 1 & 0 & 0 & 0\\
    \hline
  \end{tabular}
  \caption{Newly detected buffer overflow vulnerabilities in Android applications using PiliPidi}
  \label{tab:new_android_issues}
\end{table*}
\raggedbottom

Finally, this phase outputs potential data flow paths with incorrect buffer bounds checks as potential warnings. With this improvement also in place, the number of data flow paths with buffer bounds checks already in place was filtered as false positives from the potential number of data flow paths.

We currently handle four kinds of buffer-bound related issues. The first is when the buffer is read or written out of its bounds using index-based access. The second is when a buffer is assigned to another buffer without any check on buffer sizes for both the buffer at the left-hand and right-hand side of an assignment statement. The third is related to the second but with an additional check to see whether the assignment is properly guarded with buffer size conditional checks. The buffer size checks at the native side are usually checked with the "\textit{sizeof}" function call and then the subsequent conditional expression. Lastly, we check the buffer size and existence of conditional checks before the sinks where buffer modification happens such as calls to "\textit{memcpy}" or "\textit{strcpy}".

\section{Evaluation} \label{lab:evaluation}
To evaluate the competence of PilaiPidi in the real world, we started to look for existing benchmark suites, but unfortunately, we could not find any standard benchmark with buffer overflow vulnerabilities. Previous studies \cite{lee2020broadening}, \cite{wei2018jn} have used a custom benchmark \cite{nativeflowbench} but we did not evaluate this as it did not contain any applications with buffer overflow vulnerabilities which is our main target. The applications in this benchmark are created to show that specific sensitive information (device IMEI as a string value) is being leaked (logged) or not leaked on the native side. The studies which use this benchmark are measuring the precision and recall based on detecting or not detecting the data flow paths that actually leak the IMEI value. For our work, this is not a suitable benchmark as we want to detect apps that contain buffer overflow vulnerabilities on the native side when the data flow is originating from the Java side.

If we look at studies evaluating real-world applications, the most recent study by Lee et al. \cite{lee2020broadening} evaluated a set of Android applications, but they mainly reported findings on inter-operation issues. Since our work is on cross-language buffer overflow detection, we evaluate real-world Android projects given in Table \ref{tab:new_android_issues}. 

We used our tool on some existing well-known Android applications as given in Table \ref{tab:new_android_issues}. Our tool was able to detect 23 new buffer overflow vulnerabilities out of which, we have confirmed 11 errors with the developers of three projects (Sipdroid, camerakit-android, and mupen64plus-ae). In the same table, we can see we have evaluated a few other tools for the same Android projects. 

Flawfinder and Cppcheck are well-known static analyzers that focus only on C/C++ language. We ran these tools on the given Android projects and the results say how many warnings or issues were detected. The results for Flawfinder and Cppcheck show that they report a large number of issues or warnings, a large number of which could be false positives. But none of them were able to detect the confirmed issues detected by PilaiPidi (last row in the table). The main reason is that these tools do not have contextual information on the data flow where the source is in Java and the sink is C/C++.

We also tried to see how many sink functions exist using a simple text-based search using the grep command on the sink function list. This is given in the first row of the table. Comparing these static analysis techniques with our approach, our approach detects less number of potential issues (source-to-sink data flow paths). 

The last row from this table shows the number of confirmed issues by developers out of the exploited warnings and the row ("PilaiPidi warnings exploited by us") shows the total number of exploited warnings. The row ("PilaiPidi warnings") shows the potential warnings (data flow paths) detected by our tool where we have not exploited all of them yet. If we take Sipdroid, for example, our tool detected 19 warnings but we were able to exploit 4 paths. The remaining 15 warnings can be either potential issues that we cannot exploit yet or false positives.

The row "Precision \%" from Table \ref{tab:new_android_issues} is our precision measure by using the ``PilaiPidi warnings'' row as the total of our warnings~(true positives + false positives) and the ``PilaiPidi warnings exploited by us'' row as the true positives. Any warning that could not be exploited, we consider a false positive. While this is an extreme assumption~(just because we couldn't exploit it doesn't mean that it cannot be exploited or the developers think it cannot be exploited), we use this so that we can find a lower bound on precision. Our true precision could be higher. But we would be able to do that only by employing the developers of these apps to confirm them as false positives. Since we don't have ground truth on all possible buffer overflow vulnerabilities in the real-world dataset, we cannot measure recall.

\pagebreak
\noindent\textbf{Comparison of PilaiPidi with existing cross-language analysis tools (JN-SAF \cite{wei2018jn} and C-Summary \cite{lee2020broadening}):}

When trying to use JN-SAF tool, we faced a compilation-related issue that prevented us to build and install the tool on our machine. The issue occurs when we followed the instruction given in Argus-SAF \cite{argussaf} repository, which contains the JN-SAF and Nativedroid tools. A similar issue has been already reported in the GitHub repository but there was no official solution or fix provided yet \cite{jnsafbroken}. 

We found a docker image containing the executable JN-SAF tool, mentioned in the same GitHub issue, which we were able to set up on a docker container. We then evaluated our dataset of six Android applications on JN-SAF. But for all six applications, the tool timed out after exceeding an internal timeout of five minutes. We then tried to change this timeout value, but then figured out that it was not configurable as it was hard-coded at the code level and we are unable to compile the tool from the code due to the compilation issue mentioned above. 

Though we were not able to run the tool, we went through their code base to understand whether they can detect buffer access-related issues. The existing sinks from their code base reveal that they mainly focus on identifying system/library function invocations. So we believe that it does not have inherent support for detecting buffer access-related sinks. 

Most importantly, their current summary-based bottom approach will not work with checking for buffer access since the approach does not allow getting the value of buffer size and index size which we can get from our approach of using slice-based analysis. Without getting the values related to buffers, it will be challenging to perform a value analysis around buffer operations to detect potential buffer overflow issues.

We then tried to run the C-Summary tool on the Android projects given in Table \ref{tab:new_android_issues}, but for all six of the projects, this tool did not identify any issues based on the output we observed. We have marked this as Not detected (N/D) in the table. The main reason for this tool not being able to detect those confirmed issues is that it does not identify sinks but mainly focuses on resolving and identifying inter-operable flow-related bugs such as wrong foreign function calls and mishandling of Java exceptions. Our evaluation also shows that the C-Summary tool mainly focused on detecting inter-operation bugs.

\section{Discussion} \label{lab:discussion}
Cross-language static analysis requires precise knowledge of both source and target language syntax and semantics. There are sophisticated tools developed to do static analysis focusing on one language only \cite{ratsdocs}, \cite{wheeler2013flawfinder}, \cite{marjamaki2013cppcheck}, \cite{clanganalyzer}, \cite{brat2014ikos} and there are tools that target Android applications security from a static analysis point of view \cite{gordon2015information}, \cite{arzt2014flowdroid}, \cite{li2015iccta}, \cite{lu2012chex}. 

One can combine these tools to build a solution that can analyse both the source and target language separately and then combine the results for the purpose of cross-language analysis. In fact, some of the previous studies \cite{lee2020broadening} made use of such a technique to combine tools for different languages and provided it as the solution. Our choice of combining both languages to a common XML-based format without any loss of syntax data has shown to be effective in our case studies. 

\textbf{Choosing an XML-based AST format for analysis:}
There are other tools that can be used for multi-language code analysing. CodeQL \cite{codeql}, for instance, can be used here as it has support for representing source code for multiple languages. The main reason we did not adopt CodeQL is due to its licensing issue which conflicts with our intention in shipping the library as a plugin for a commercial IDE. Additionally, we believe that our approach of using forward source slicing can be used on top of common AST representation that can be produced for multiple languages (especially Java and C/C++) by any tool. Also, with a common XML-based AST format, the underlying logic to read, parse, and analyse the XML format will not change, when expanding to other language pairs, which we believe is the major advantage over previous work in this area.

\textbf{Orthogonal to existing approaches:}
Though our attempts with evaluating the tools from previous work \cite{lee2020broadening}, \cite{wei2018jn} were unsuccessful, we believe they are capable of detecting bugs on their own merits. C-Summary \cite{lee2020broadening} tool is capable of resolving bi-directional inter-operation which they claim to outperform JN-SAF \cite{wei2018jn} in their evaluation. Whereas JN-SAF, which is based on binary code analysis has evaluated existing malware applications from a known dataset and was able to detect potential malware. In addition to that, with binary code-based analysis, we believe they mainly target platform integration such as Google Play Store, where they could run their tool by providing the applications APK. 

Our approach, on the other hand, is orthogonal to the previous studies~\cite{lee2020broadening}, \cite{wei2018jn} and relies on the source code of the application. By using source code, our analysis is able to do things like precise bound checks that other techniques cannot. Our main target is integration with the development stage and detecting bugs as early as possible before pushing the application into production.

\textbf{Detecting context sensitive buffer overflow:}
Buffer overflow is one of the most occurring ones in Android applications according to the study from Aloraini and Nagappan \cite{aloraini2017evaluating}. The study \cite{aloraini2017evaluating} shows that existing state-of-the-art tools were not able to identify cross-language buffer overflow vulnerabilities. The main reasons are that existing tools produce too many false positives and they are not capable of detecting data flows originating from the source language Java and ending in the target language C/C++. 

Detecting buffer overflow in the target language still remains a challenge, but with the use of PilaiPidi, we can detect cross-language buffer overflow. As a start, we have shown that PilaiPidi is capable of detecting new previously undetected buffer overflow vulnerabilities. We also believe that the data flow analysis of PilaiPidi could be used as the base for many types of issues that can potentially be detected during compile time such as memory corruption errors and cross-site scripting errors which can occur at the native side of the program. The current implementation of PilaiPidi can be extended to do this analysis.

\textbf{Unexploited warnings from PilaiPidi:}
From Table \ref{tab:new_android_issues}, we can see that there are still more warnings reported from PilaiPidi than exploited warnings. The exploited warnings are the ones we were able to exploit until this paper is written. We are still working on how to properly exploit the remaining warnings. We believe that some of the warnings could be false positives which our tool cannot detect based on the current approach. 

We currently only check the buffer access-related validation at the C/C++ side within the scope of the enclosing function where the buffer access sink is identified. As the data flow starts from the Java side, and the input from the Java side is already validated, then any data flow paths from that location will not exploitable. This is a limitation for our work as we only check the validation on the C/C++ side. But we believe if we extend the validation check along the data flow path from sink to source, we should be able to further reduce the false positives. But as of now, we have not ruled out any of the reported warnings as false positives as we continue to exploit the paths reported from the warnings.

\section{Related Work} \label{lab:related_work}
Previous studies have looked at program analysis including cross-language analysis in two broad areas, namely static and dynamic analysis. In static analysis-based approaches, one line of work focuses on analysing the program at compile time using source code analysis, and another line of work focuses on compiled code (byte code or binary) level analysis. The dynamic analysis mainly focuses on analyses by executing the program. For this work, we focus on only the static analysis approaches. 

Static analysis-based solutions of programs, with single or multiple languages, perform either source code analysis or byte code and binary code analysis. In most cases, the code is statically transformed to another representation to perform the analysis easily. There is a line of work that focuses on Android application security issues but the majority of the work only focuses on the Java side of the application while few studies have looked at addressing security-related issues on the native side as well.

\subsection{Source code based analysis}
The work from Lee et al. \cite{lee2020broadening} is the most recent study which proposes a source code-based static analysis technique. Using modular analysis, they extract semantic summary from the C language and utilize a host language-specific static analyser and combine both results to build the complete call graph. It is built on top of Amandroid \cite{wei2018amandroid}. The main difference between this study and ours is that it utilizes separate static analysers of a host (Java) and a guest (C) languages separately while our work does not distinguish between host and guest languages. Also, the study from Lee et al. \cite{lee2020broadening}, to our understanding, focuses mainly on detecting inter-operation bugs, such as missing functions, wrong exception handling, etc. by building and analysing call graphs. But in our work, we have focused on detecting security issues (buffer overflow), which occur on the native side with data flow analysis.

\subsection{Byte code or binary code based analysis}
Joern \cite{joernbughunter} generates code property graphs to represent the code and does analysis on the binary code. The main difference is that we are in the static analysis domain with information from source code, whereas Joern relies on binary code information with specific tools underneath to support different languages with its platform.

JN-SAF \cite{wei2018jn} is capable of performing cross-language analysis using DEXtoIR to convert Android byte code to an intermediate representation, translate the binary code to another intermediate representation, build call graphs, generate a unified heap manipulation summary for both Java and native functions and finally performs symbolic execution at the native binary code level. But with generating heap summaries of the compiled code, JN-SAF cannot do buffer size value analysis.

Amandroid \cite{wei2018amandroid} is a data flow analysis framework, which generates an environment model for each Android component and applies a component-based analysis algorithm to capture inter-component data flows at the Java layer. FlowDroid \cite{arzt2014flowdroid} performs static taint analysis for Android applications. It properly handles Android framework's life-cycle events and callbacks by modeling them and performing flow and context-sensitive analysis to detect taint propagation at the Java layer.

DroidSafe \cite{gordon2015information} tracks data flow in RPC calls and Intents in Android applications and handles run time event ordering by precisely modeling combining it with static analysis. Epicc \cite{octeau2013effective} extends FlowDroid to effectively find inter-component communication and call parameters of Intents in the Android environment. IccTA \cite{li2015iccta}, also extends FlowDroid to detect privacy leaks in inter-component communication and tracks tainted data flow by leveraging Android Intent call and return information.

CHEX \cite{lu2012chex} performs low-overhead reachability analysis on dependency graphs and is designed to detect component hijacking-related security issues in Android. It used Wala \cite{dolby2015tj} underneath to build the reachability graph by generating app splits and building a data flow summary. As we can see the above tools operate only at the Java layer of the applications.

The tool \textit{angr}\cite{shoshitaishvili2016state} supports different system architectures and it is a framework for analysing binaries and combines both static and dynamic symbolic analysis by first translating assembly code to an IR and then performing symbolic execution on it. BitBlaze \cite{song2008bitblaze} is a hybrid binary analysis platform that first translates binary code to an IR supporting x86 and ARMv4 architecture and then instruments the binary to finally perform symbolic execution. 

These binary code analysis solutions are specifically developed to analyse assembly code, which requires the program to be compiled first. But PilaiPidi does not require code to be compiled first and it operates on top of the source code of the applications. Overall, as we can see from the above-related work, the majority of them only focus on the Java side while ignoring or not handling the native method invocations and another line of work only focused on native code only. But our work, on the other hand, performs cross-language data flow analysis and also has the potential to expand analysing of many source and target language pairs.

\section{Conclusion} \label{lab:conclusion}
In this paper, we presented a cross-language static data flow analysis tool to detect buffer overflow vulnerabilities in applications written in Java and C/C++. Our open-source tool, PilaiPidi \cite{pilaipidi}, performs a forward program slice-based technique to build and analyse data flow across the source and target language. Our experiment and evaluation results showed that PilaiPidi is capable of detecting new previously unknown buffer-related issues in real-world Android applications. In addition to this, the data flow analysis of PilaiPidi can potentially be the basis for any other analysis to detect various types of cross-language bugs and issues.

\section{Data Availability} The code for PilaiPidi and the source code for the open-source Android apps that we used in our evaluation have been uploaded to a public repository~\cite{pilaipididataset}. 

\bibliographystyle{acm}
\bibliography{sample}

\begin{thebibliography}{10}

\bibitem{argussaf}
{Argus-SAF: Argus static analysis framework}.
\newblock \url{https://github.com/arguslab/Argus-SAF}.
\newblock [Online; accessed 01-February-2023].

\bibitem{asciicamfixes}
{AsciiCam fixes}.
\newblock \url{https://github.com/dozingcat/AsciiCam/pull/26}.
\newblock [Online; accessed 01-February-2023].

\bibitem{camarakitfixes}
{Camerakit Android fixes}.
\newblock \url{https://github.com/CameraKit/camerakit-android/pull/635}.
\newblock [Online; accessed 01-February-2023].

\bibitem{clanganalyzer}
{Clang Static Analyzer}.
\newblock \url{https://clang-analyzer.llvm.org/}.
\newblock [Online; accessed 01-February-2023].

\bibitem{codeql}
{CodeQL - A semantic code analysis engine}.
\newblock \url{https://codeql.github.com/}.
\newblock [Online; accessed 01-February-2023].

\bibitem{pilaipididataset}
{Dataset for PilaiPidi - Collection of Android applications used for
  evaluation}.
\newblock \url{https://figshare.com/s/98a23508e1bbb56e4d50}.
\newblock [Online; accessed 01-February-2023].

\bibitem{frozenbubblefixes}
{Frozenbubble fixes}.
\newblock \url{https://github.com/videogameboy76/frozenbubbleandroid/pull/65}.
\newblock [Online; accessed 01-February-2023].

\bibitem{jnsafbroken}
{Jnsaf and nativedroid installation failed - GitHub Issue}.
\newblock \url{https://github.com/arguslab/Argus-SAF/issues/83}.
\newblock [Online; accessed 01-February-2023].

\bibitem{joernbughunter}
{Joern - The Bug Hunter's Workbench}.
\newblock \url{https://github.com/joernio/joern}.
\newblock [Online; accessed 01-February-2023].

\bibitem{mupen64plusfixes}
{Mupen64Plus-AE fixes}.
\newblock \url{https://github.com/mupen64plus-ae/mupen64plus-ae/pull/1039}.
\newblock [Online; accessed 01-February-2023].

\bibitem{nativeflowbench}
{NativeFlowBench}.
\newblock \url{https://github.com/arguslab/NativeFlowBench}.
\newblock [Online; accessed 01-February-2023].

\bibitem{pilaipidi}
{PilaiPidi - A Cross-language data flow analyser}.
\newblock \url{https://figshare.com/s/b51ed2b32a8bb702dd95}.
\newblock [Online; accessed 01-February-2023].

\bibitem{pilaipidifixes}
{Pull requests for the exploited warnings from the first author's GitHub
  account - withheld for double-blind submission sake}.
\newblock [Online; accessed 01-February-2023].

\bibitem{ratsdocs}
{Rough Auditing Tool for Security (RATS)}.
\newblock
  \url{https://code.google.com/archive/p/rough-auditing-tool-for-security/}.
\newblock [Online; accessed 01-February-2023].

\bibitem{sibdroidfixes}
{Sipdroid fixes}.
\newblock \url{https://github.com/i-p-tel/sipdroid/blob/wiki/ChangeLog.md}.
\newblock [Online; accessed 01-February-2023].

\bibitem{srcmldoc}
{srcML Documentation}.
\newblock \url{https://www.srcml.org/doc/cpp\_srcML.html}.
\newblock [Online; accessed 01-February-2023].

\bibitem{wonderdroidfixes}
{Wonderdroid-X fixes}.
\newblock \url{https://github.com/williehwc/wonderdroid-x/pull/16}.
\newblock [Online; accessed 01-February-2023].

\bibitem{aloraini2017evaluating}
{\sc Aloraini, B., and Nagappan, M.}
\newblock Evaluating state-of-the-art free and open source static analysis
  tools against buffer errors in android apps.
\newblock In {\em 2017 IEEE International Conference on Software Maintenance
  and Evolution (ICSME)\/} (2017), IEEE, pp.~295--306.

\bibitem{arzt2013susi}
{\sc Arzt, S., Rasthofer, S., and Bodden, E.}
\newblock Susi: A tool for the fully automated classification and
  categorization of android sources and sinks.
\newblock {\em University of Darmstadt, Tech. Rep. TUDCS-2013-0114\/} (2013).

\bibitem{arzt2014flowdroid}
{\sc Arzt, S., Rasthofer, S., Fritz, C., Bodden, E., Bartel, A., Klein, J.,
  Le~Traon, Y., Octeau, D., and McDaniel, P.}
\newblock Flowdroid: Precise context, flow, field, object-sensitive and
  lifecycle-aware taint analysis for android apps.
\newblock {\em Acm Sigplan Notices 49}, 6 (2014), 259--269.

\bibitem{brat2014ikos}
{\sc Brat, G., Navas, J.~A., Shi, N., and Venet, A.}
\newblock Ikos: A framework for static analysis based on abstract
  interpretation.
\newblock In {\em International Conference on Software Engineering and Formal
  Methods\/} (2014), Springer, pp.~271--277.

\bibitem{collard2013srcml}
{\sc Collard, M.~L., Decker, M.~J., and Maletic, J.~I.}
\newblock srcml: An infrastructure for the exploration, analysis, and
  manipulation of source code: A tool demonstration.
\newblock In {\em 2013 IEEE International Conference on Software Maintenance\/}
  (2013), IEEE, pp.~516--519.

\bibitem{Ding_2014}
{\sc Ding, S., Tan, H. B.~K., and Zhang, H.}
\newblock Abor: An automatic framework for buffer overflow removal in c/c++
  programs.
\newblock 204--221.

\bibitem{dolby2015tj}
{\sc Dolby, J., Fink, S.~J., and Sridharan, M.}
\newblock {TJ Watson libraries for analysis (WALA)}.
\newblock {\em URL http://wala.sf.net\/} (2015).

\bibitem{gordon2015information}
{\sc Gordon, M.~I., Kim, D., Perkins, J.~H., Gilham, L., Nguyen, N., and
  Rinard, M.~C.}
\newblock Information flow analysis of android applications in droidsafe.
\newblock In {\em NDSS\/} (2015), vol.~15, p.~110.

\bibitem{gordon1998essential}
{\sc Gordon, R.}
\newblock {\em Essential JNI: Java Native Interface}.
\newblock Prentice-Hall, Inc., 1998.

\bibitem{lee2020broadening}
{\sc Lee, S., Lee, H., and Ryu, S.}
\newblock Broadening horizons of multilingual static analysis: semantic summary
  extraction from c code for jni program analysis.
\newblock In {\em Proceedings of the 35th IEEE/ACM International Conference on
  Automated Software Engineering\/} (2020), pp.~127--137.

\bibitem{Lhee_2003}
{\sc Lhee, K.-S., and Chapin, S.~J.}
\newblock Buffer overflow and format string overflow vulnerabilities.
\newblock {\em Softw. Pract. Exper. 33}, 5 (apr 2003), 423–460.

\bibitem{li2015iccta}
{\sc Li, L., Bartel, A., Bissyand{\'e}, T.~F., Klein, J., Le~Traon, Y., Arzt,
  S., Rasthofer, S., Bodden, E., Octeau, D., and McDaniel, P.}
\newblock Iccta: Detecting inter-component privacy leaks in android apps.
\newblock In {\em 2015 IEEE/ACM 37th IEEE International Conference on Software
  Engineering\/} (2015), vol.~1, IEEE, pp.~280--291.

\bibitem{lu2012chex}
{\sc Lu, L., Li, Z., Wu, Z., Lee, W., and Jiang, G.}
\newblock Chex: statically vetting android apps for component hijacking
  vulnerabilities.
\newblock In {\em Proceedings of the 2012 ACM conference on Computer and
  communications security\/} (2012), pp.~229--240.

\bibitem{marjamaki2013cppcheck}
{\sc Marjam{\"a}ki, D.}
\newblock {Cppcheck: a tool for static C/C++ code analysis}.
\newblock \url{http://cppcheck.sourceforge.net/}, 2013.

\bibitem{newman2016srcslice}
{\sc Newman, C.~D., Sage, T., Collard, M.~L., Alomari, H.~W., and Maletic,
  J.~I.}
\newblock srcslice: A tool for efficient static forward slicing.
\newblock In {\em 2016 IEEE/ACM 38th International Conference on Software
  Engineering Companion (ICSE-C)\/} (2016), IEEE, pp.~621--624.

\bibitem{octeau2013effective}
{\sc Octeau, D., McDaniel, P., Jha, S., Bartel, A., Bodden, E., Klein, J., and
  Le~Traon, Y.}
\newblock Effective $\{$Inter-Component$\}$ communication mapping in android:
  An essential step towards holistic security analysis.
\newblock In {\em 22nd USENIX Security Symposium (USENIX Security 13)\/}
  (2013), pp.~543--558.

\bibitem{ratabouil2015android}
{\sc Ratabouil, S.}
\newblock {\em Android NDK: beginner's guide}.
\newblock Packt Publishing Ltd, 2015.

\bibitem{shoshitaishvili2016state}
{\sc Shoshitaishvili, Y., Wang, R., Salls, C., Stephens, N., Polino, M.,
  Dutcher, A., Grosen, J., Feng, S., Hauser, C., Kruegel, C., and Vigna, G.}
\newblock {SoK: (State of) The Art of War: Offensive Techniques in Binary
  Analysis}.
\newblock In {\em IEEE Symposium on Security and Privacy\/} (2016).

\bibitem{song2008bitblaze}
{\sc Song, D., Brumley, D., Yin, H., Caballero, J., Jager, I., Kang, M.~G.,
  Liang, Z., Newsome, J., Poosankam, P., and Saxena, P.}
\newblock Bitblaze: A new approach to computer security via binary analysis.
\newblock In {\em International conference on information systems security\/}
  (2008), Springer, pp.~1--25.

\bibitem{wei2018jn}
{\sc Wei, F., Lin, X., Ou, X., Chen, T., and Zhang, X.}
\newblock Jn-saf: Precise and efficient ndk/jni-aware inter-language static
  analysis framework for security vetting of android applications with native
  code.
\newblock In {\em Proceedings of the 2018 ACM SIGSAC Conference on Computer and
  Communications Security\/} (2018), pp.~1137--1150.

\bibitem{wei2018amandroid}
{\sc Wei, F., Roy, S., and Ou, X.}
\newblock Amandroid: A precise and general inter-component data flow analysis
  framework for security vetting of android apps.
\newblock {\em ACM Transactions on Privacy and Security (TOPS) 21}, 3 (2018),
  1--32.

\bibitem{wheeler2013flawfinder}
{\sc Wheeler, D.~A.}
\newblock {Flawfinder}.
\newblock \url{http://www.dwheeler.com/flawfinder/}, 2013.

\end{thebibliography}




\end{document}